\documentclass[twocolumn,preprintnumbers,amsmath,amssymb,floatfix,showpacs,prl]{revtex4}

\usepackage{dcolumn}
\usepackage{bm}
\usepackage{graphicx}

\begin{document}


\title{Geometrical Description of the  Fractional Quantum Hall Effect}

\author{F. D. M. Haldane}
\affiliation{Department of Physics, Princeton University,
Princeton NJ 08544-0708}

\date{June 16, 2011}

\begin{abstract}
The fundamental collective degree of freedom of 
fractional quantum Hall states is identified as a unimodular
two-dimensional spatial metric that characterizes the local 
shape of the correlations of the incompressible fluid. Its
quantum fluctuations are controlled by a topologically-quantized
``guiding-center spin''.  Charge fluctuations are proportional to its
Gaussian curvature.
\end{abstract}
\maketitle

In this Letter, I point out the apparently previously-unnoticed
geometric degree of freedom of the fractional quantum Hall effect (FQHE),
that fundamentally distinguishes it from the integer effect, and will
provide the basis for a new  description of its collective
properties as a fluctuating quantum geometry.

The simplest model Hamiltonian for $N$ interacting electrons bound to a
 two-dimensional (2D) planar ``Hall
surface'' traversed by a uniform magnetic flux density 
is 
\begin{equation}
H = \sum_{i=1}^N \frac{1}{2m}g^{ab}\pi_{ia}\pi_{ib} +
\frac{1}{A}\sum_{\bm q} V(\bm q) \sum_{i<j} e^{i\bm q\cdot (\bm r_i-\bm
  r_j)}. 
\end{equation}
Here $\bm r_i - \bm r_j$ = $(r^a_i-r^a_j)\bm e_a$,  $[r^a_i,r^b_j]$ = 0,
are the relative
displacements
of the particles on the 2D surface
 with orthonormal tangent vectors $\bm e_a$, $a=1,2$, and
 $\pi_{ia}$ = $\bm e_a\cdot \bm \pi_i$
are the components of the gauge-invariant
dynamical momenta, with commutation relations
\begin{equation}
[r_i^a,\pi_{jb}]= i\delta_{ij}\hbar \delta^a_b, \quad 
[\pi_{ia},\pi_{jb}] = i\delta_{ij}\epsilon_{ab}\hbar^2/\ell_B^2.
\end{equation}
I use Einstein summation convention: $q_ar^a$ = $\bm q\cdot \bm r$
(index placement distinguishes real-space vectors $r^a$ from dual
(reciprocal)
vectors $q_a$); $\delta^a_b$ is the Kronecker symbol, and
$\epsilon_{ab}$ = $\epsilon^{ab}$ is the 2D antisymmetric Levi-Civita
symbol.
A periodic boundary condition (pbc) can be imposed on a fundamental region of
the plane with area $A$ = $2\pi\ell^2_BN_{\Phi}$,
which restricts wavevectors $\bm q$ to the reciprocal
lattice; $N_{\Phi}$ is an integer, and $2\pi \ell_B^2$ is the area
through which a London magnetic flux quantum $h/e$ passes.

The parameters of the Hamiltonian are: (1) a Galileian effective mass
tensor $mg_{ab}$, where $g_{ab}$ is a 
positive-definite ``Galileian metric''
 with $\det g$ = 1 (\textit{i.e.}, a
\textit{unimodular} metric) 
and inverse $g^{ab}$, and 
$m > 0$ is the effective mass that controls the cyclotron frequency
$\omega_B$ = $\hbar/m\ell_B^2$; (2) $V(\bm q)$ which is the
Fourier transform of an unretarded translationally-invariant
two-body interaction potential.    

In principle, the real function $V(\bm q)$ is the
Fourier transform of the long-ranged unscreened  Coulomb potential,
with the small-$\bm q$ behavior
\begin{equation}
\lim_{\lambda \rightarrow 0} \lambda V(\lambda \bm q) \rightarrow 
\frac{e^2}{2\varepsilon}(\tilde g^{ab}q_aq_b)^{-1/2} ,
\end{equation} 
where $\tilde g^{ab}$ is the inverse of a unimodular \textit{Coulomb
  metric} $\tilde g_{ab}$, controlled by the dielectric properties of
the surrounding 3D insulating media, while the large-$\bm q$ behavior
of
$V(\bm q)$ is controlled by  the quantum well that binds
electrons to the surface.  The singularity of  $V(0)$ does not affect
incompressibility, and can be screened by
a metallic plane placed parallel to the  surface.

There is no fundamental reason for the Coulomb and Galileian
metrics to coincide, unless  there is an atomic-scale discrete
$(n>2)$-fold rotational symmetry of the surface, and no
tangential magnetic flux.    I will argue that the usual implicit assumption of
 rotational symmetry hides
key geometric features of the FQHE.

In the presence of the magnetic field, the canonical degrees of
freedom $\{\bm r_i, \bm p_i\}$ are reorganized into two independent sets,
the dynamical momenta $\{\bm \pi_i\}$, which I will call ``left-handed'' 
degrees of freedom, and the ``guiding centers'' $\{ \bm
R_i\}$, the ``right-handed''  degrees of freedom,
\begin{equation}
 R_i^a = r_i^a -\hbar^{-1}\epsilon^{ab}\pi_{ib}\ell_B^2, \quad
[R^a_i,R^b_j] = -i\delta_{ij}\epsilon^{ab}\ell_B^2,
\label{qgeom}
\end{equation}
with $[R^a_i,\pi_{jb}]$ = 0.
The pbc further restricts the guiding-center variables
to the set of unitary operators
$\rho_{\bm q,i}$ = $\exp {i\bm q\cdot \bm R_i}$,
which obey the Heisenberg algebra
\begin{equation}
\rho_{\bm q,i}\rho_{\bm q',i} = e^{i\frac{1}{2}\bm q\times \bm
  q'\ell_B^2}\rho_{\bm q + \bm q',i}, \quad \bm q \times \bm q' \equiv \epsilon^{ab}q_aq'_b;
\end{equation}
reciprocal vectors $\bm q, \bm q'$ compatible with the
pbc obey
$( \exp {i\bm q\times \bm q'\ell_B^2} )^{N_{\Phi}}$ = 1.
The pbc can be expressed as 
\begin{equation}
\left (\rho_{\bm q,i}\right )^{N_{\Phi}}|\Psi\rangle  = \left
  (\eta_{\bm q}\right )^{N_{\Phi}}|\Psi \rangle
\end{equation}
for all states in the Hilbert space, where $\eta_{\bm q} = 1$ if
$\frac{1}{2}\bm q$ is an allowed reciprocal vector, and $\eta_{\bm q}
= -1$ otherwise.  
This leads to the recurrence relation
\begin{equation}
\rho_{\bm q + N_{\phi}\bm q',i} = \left (\eta_{\bm q'}e^{i\frac{1}{2}\bm q \times
    \bm q'\ell_B^2}\right )^{N_{\Phi}}\rho_{\bm q,i} = \pm \rho_{\bm
    q,i}.
\end{equation}
For a given particle label $i$,  the set of
independent operators $\bm \rho_{\bm q,i}$ can be reduced to  a set of
$N_{\Phi}^2$ operators where $\bm q \in \text{BZ}$ takes one of a set of
$N_{\Phi}^2$ distinct values that define an analog of a ``Brillouin
zone''.  Let
\begin{equation}
\delta^2_{\bm q,\bm q'} \equiv  \frac{1}{N_{\Phi}}{\sum_{ \bm q''}}' e^{i\bm
  q''\times (\bm q - \bm q')}.
\end{equation}
(Primed sums are over the BZ.)
Then $\delta^2_{\bm q,\bm q'} = 0$ if $\bm q $ and $\bm q'$ are
distinct, and has the value $N_{\Phi}$ if they are equivalent; with
this  definition $\delta^2_{\bm q,\bm q'}$ becomes $2\pi \delta^2(\bm
q\ell_B-\bm q'\ell_B)$  in the limit $N_{\Phi}\rightarrow \infty$, where
$\delta^2(\bm x)$ is the 2D Dirac delta-function.  It is convenient
to choose the BZ so it has inversion
symmetry: $\bm q \in \text{BZ}$ $\rightarrow$  $-\bm q \in \text{BZ}$,
and $\rho_{\bm q=0,i}$ is the identity.  The set of $N_{\Phi}^2-1$
operators  $\{\rho_{\bm q,i},\bm q \in \text {BZ}, \bm q \ne 0\}$ are
the generators of the Lie algebra $SU(N_{\Phi})$.
Both $\rho_{\bm q,i}$ and also (as noted by Girvin, MacDonald and
Platzman\cite{GMP}) the ``coproduct'' $\rho_{\bm q}$ =
$\sum_i \rho_{\bm q,i}$, obey 
\begin{equation}
[\rho_{\bm q},\rho_{\bm q'} ] = 2i \sin ( {\textstyle\frac{1}{2}}\bm q\times \bm
q'\ell_B^2)\rho_{\bm q+\bm q'}.
\end{equation}
In this form of the  Lie algebra, the quadratic Casimir is
\begin{equation}
C_2 = \frac{1}{2N_{\Phi}}{\sum_{\bm q\ne 0}}'
 \rho_{\bm q}\rho_{-\bm   q} = \frac{N(N_{\Phi}^2 -N)}{2N_{\Phi}} + \sum_{i<j}P_{ij},
\end{equation}
where $P_{ij}$ exchanges guiding centers of particles $i$ and $j$.
For $N=1$, the  $\rho_{\bm q,i}$ form the
$N_{\Phi}$-dimensional fundamental (defining) $SU(N_{\Phi})$ representation of
one-particle
states of a Landau level, with
 $C_2$ = $(N_{\Phi}^2 -1)/2N_{\Phi}$.

The high-field condition  is defined by
\begin{equation}
\hbar \omega_B \gg \frac{1}{A}\sum_{\bm q} V(\bm q) f(\bm q)^2,
\quad f(\bm q) = e^{-\frac{1}{4}q_g^2\ell_B^2 },
\end{equation} 
where $f(\bm q)$ is the  lowest-Landau-level  form-factor, and
$q_g^2$ $\equiv$ $ g^{ab}q_aq_b$. 
In this limit, the low-energy eigenstates of the model have all the
particles  in the lowest Landau level, and can be factorized into 
a simple \textit{unentangled product} of  states of  right-handed and left-handed degrees of
freedom:
\begin{equation}
|\Psi_{0,\alpha}\rangle =  |\Psi_0^L(g)\rangle \otimes |\Psi^R_{\alpha}\rangle,
\end{equation}
where $|\Psi^L_0(g)\rangle$ is a trivial harmonic-oscillator coherent
state,
fully symmetric under interchange of the dynamical momenta of any pair
of particles, and
parametrized only by the Galileian metric $g_{ab}$; its defining
property is
\begin{equation}
a_i |\Psi^L_0( g)\rangle = 0, 
\quad a_i \propto \omega^a(g)\pi_{ia},\quad i = 1,\ldots,N,
\label{lll}
\end{equation}
where the complex unit vector $\omega^a(g)$ is obtained by solution
of the generalized Hermitian eigenvector problem
\begin{equation}
\omega_a(g) = g_{ab}\omega^b(g) = i\epsilon_{ab}\omega^b(g), \quad
\omega_a(g)^*\omega^a(g) = 1.
\end{equation}
In contrast, the non-trivial states $|\Psi^R_{\alpha}\rangle$ are the
eigenstates of the ``right-handed'' (guiding-center) Hamiltonian
\begin{equation}
H_R = \frac{1}{2A}\sum_{\bm q}V(\bm q)f(\bm q)^2\rho_{\bm q}\rho_{-\bm q}.
\label{hamR}
\end{equation}

The reduction of the problem by discarding  ``left-handed'' degrees
of freedom,  ``frozen out'' by Landau quantization,  makes
numerical study of the problem by exact diagonalization of $H_R$ for
finite $N, N_{\Phi}$ tractable.  This may also be characterized
as a ``quantum geometry'' description: once the
``left-handed'' degrees of freedom are removed,  the notion of
\textit{locality}, fundamental to both classical geometry and Schr\"odinger's
formulation
of quantum mechanics, is absent.    The commutation relations
(\ref{qgeom}) imply a fundamental uncertainty  in the ``position'' of
the particles, now only described by their guiding centers.   A
Schr\"{o}dinger wavefunction can only be constructed after ``gluing''
$|\Psi^R_{\alpha}\rangle$ together with some $|\Psi^L\rangle$, after
which the composite state can be projected onto simultaneous eigenstates of the
commuting set $\{\bm r_i\}$: \textit{e.g.},
\begin{equation}
\Psi_{\alpha}(\{\bm r_i\},g) = \langle \{\bm r_i\} 
 |\Psi^L_0(g)\rangle \otimes |\Psi^R_{\alpha}\rangle .
\label{wf}
\end{equation}
Note that the construction (\ref{wf}) of a
Schr\"odinger wavefunction involves an \textit{extraneous quantity $(g_{ab})$
that is not directly determined by $|\Psi^R_{\alpha}\rangle$ itself},
and thus is a non-primitive construction that does not
represent
$|\Psi^R_{\alpha}\rangle$ in its purest form.
This suggests a reconsideration of the meaning of the
``Laughlin state'', usually presented in the form of the ``Laughlin
wavefunction''\cite{laughlin83}, which is fundamental to current understanding of the FQHE.

The conventional presentation of 
FQHE states  is as an $N$-particle  Schr\"odinger wavefunction with the 
 form
\begin{equation}
\Psi(\{\bm r_i\}) =
F(\{z_i\})\prod_{i=1}^Ne^{-\frac{1}{2}z_i^*z_i},
\label{llwf}
\end{equation}
where   $z_i$ = $\omega_a(g)r^a_i/\surd 2 \ell_B$.
Such wavefunctions, formulated in the ``symmetric gauge'', 
obey (\ref{lll}) with
$a_i$  $\equiv$  ${\textstyle \frac{1}{2}}z_i +
  {\partial}/{\partial z_i^*}$.
The original  Laughlin wavefunction\cite{laughlin83} was the polynomial
\begin{equation}
F(\{z_i\}) = F_L^q(\{z_i\}) \equiv  \prod_{i>j}(z_i-z_j)^q;
\label{polyl}
\end{equation}
it was subsequently adapted\cite{halrez85} to a pbc with the form
\begin{equation}
F^q_{L,\alpha}(\{z_i\}) = \prod_{i>j}w(z_i-z_j)^q \prod_{k=1}^q
w(({\textstyle \sum _i} z_i)-a_{k,\alpha}),
\label{lwfpbc}
\end{equation}
where $w(z)$ is given in terms of an elliptic
theta function:
$w(z)$ = $\theta_1 (\pi z/L_1|L_2/L_1)\exp (z^2/2N_{\Phi} )$,
with
$L_1L_2^* - L_1^*L_2$ = $2\pi i N_{\Phi}$ (the wavefunction is (quasi)
periodic under $z_i$ $\rightarrow$ $z_i + mL_1 + nL_2$). 
The additional $q$ parameters $a_{k,\alpha}$ of (\ref{lwfpbc}), with $\sum_k a_{k,\alpha}$ = 0, characterize the
$q$-fold topological degeneracy  of the  Laughlin state with a pbc.

The Laughlin wavefunction was originally presented as a ``variational
wavefunction'', albeit one with no continuously-tunable parameter, since
$q$ is an integer fixed by statistics.  Its initial success
was that, as a ``trial wavefunction'', it had a lower Coulomb
energy than obtained in Hartree-Fock approximations, and
explained the existence of  a strong FQHE state at $\nu$
$\equiv$ $N/N_{\Phi}$
= 1/3, but not at $\nu$ = 1/2.    In the wavefunction language, its
defining
characteristic is that, as a function of any particle coordinate
$z_i$, there is an order-$q$ zero at the location of every other
particle, which heuristically ``keeps particles apart'', and lowers the Coulomb 
energy.   

Subsequent to  its introduction, the Laughlin state's essential
validity  was further confirmed
by this author's observation\cite{haldane83} that, at $\nu = 1/q$, it is also uniquely characterized as
the  zero-energy eigenstate of a  two-body ``pseudopotential Hamiltonian''
\begin{equation}
H_R = \sum_{m=0}^{q-1}V_mP_m(g) ,\quad V_m > 0,
\end{equation}
where
\begin{equation}
P_m(g) = \frac{1}{N_{\Phi}}\sum_{\bm q}
  L_m(q_g^2\ell_B^2)e^{-{\textstyle\frac{1}{2}}  q^2_g\ell_B^2}
\rho_{\bm q}\rho_{-\bm q},
\end{equation}
where $L_m(x)$ is a Laguerre polynomial.
Numerical finite-size diagonalization\cite{halrez85a} for $q$ = 3 showed that this $H_R$
had the gapped excitation spectrum of an incompressible FQHE state,
and that this gap  did not close along a path that adiabatically
interpolated between it and the Hamiltonian of the Coulomb
interaction
with $\tilde g_{ab}$ = $g_{ab}$.

This raises the question that does not seem to have been previously
considered: what if  the ``Coulomb metric''
$\tilde g_{ab}$ and the ``Galileian metric'' $g_{ab}$ do \textit{not} coincide?
The ``pseudopotential'' definition of the Laughlin \textit{state} (as opposed
to the Laughlin \textit{wavefunction}) defines a
\textit{continuously-parametrized family}
of $\nu$ = $1/q$  Laughlin states $|\Psi^q_{L,\alpha} (\bar g)\rangle$ by
\begin{equation}
P_m(\bar g)|\Psi^q_{L,\alpha}(\bar g)\rangle = 0 , m < q.
\end{equation} 
The continuously-variable parameter here is a unimodular \textit{guiding-center
  metric} $\bar g_{ab}$ that is in principle distinct from the
Galileian metric $g_{ab}$, and is \textit{not} fixed by the one-body physics of the
Landau orbits.   Physically,  it characterizes the \textit{shape} of the
correlation functions of the Laughlin state.  If the shape of  Landau orbits is
used as the definition of  ``circular'', the correlation hole around
the particles  deforms to ``elliptical'' when $\bar g_{ab}$ $\ne $
$g_{ab}$.

 If a wavefunction (\ref{lll}) is constructed by
``gluing together'' $|\Psi^L_0( g)\rangle $ with the ``Laughlin
\textit{state}'' $|\Psi^R\rangle$ =
$|\Psi^q_{L,\alpha}(\bar g)\rangle$,
it does \textit{not} correspond to the Laughlin \textit{wavefunction} (\ref{lwfpbc})
\textit{unless} $\bar g_{ab}$ = $g_{ab}$, as  there is no longer a
$q$'th order zero of the wavefunction when $z_i$ = $z_j$.
Despite this, I will not
call  $|\Psi^q_L(\bar g)\rangle$ with $\bar g_{ab}$  $\ne$ $g_{ab}$ 
a ``generalization'' of the
Laughlin state, but propose it as a 
definition of the \textit{family} of  Laughlin states
that exposes the  geometrical
degree of  freedom $\bar g_{ab}$ hidden by the
wavefunction-based formalism.   I argue that FQHE states
should be described completely within the framework of the
``quantum geometry'' of the guiding-center degrees of
freedom alone, and no ``preferred status'' should be accorded to the metric
choice $\bar g_{ab}$ = $g_{ab}$.
If the states $|\Psi_{L,\alpha}^q(\bar g)\rangle$ are used as
variational approximations to the ground state of a generic $H_R$
given by (\ref{hamR}), $\bar g_{ab}$ must be chosen to minimize the
correlation energy $E(\bar g)$ = $\langle \Psi_{L\alpha}^q(\bar g)|H_R
|\Psi_{L\alpha}^q(\bar g)\rangle$.   If the
Coulomb ($\tilde g_{ab}$) and Galileian ($g_{ab}$)  metrics coincide, the energy will
be minimized by the choice $\bar g_{ab}$ = 
$\tilde g_{ab}$= $g_{ab}$; otherwise, $\bar g_{ab}$ will be a
compromise intermediate between the two physical metrics.

A more profound consequence of the identification of the variable geometric
parameter $\bar g_{ab}$
follows from the observation that the correlation energy 
will be a quadratic function of local deformations $\bar g_{ab}(\bm r,t)$ around
the minimizing value, whether or
not this is equal to $g_{ab}$. This unimodular metric, or ``shape of the circle''
defined by the correlation function of the FQHE state, may be
identified as the natural \textit{local collective degree of freedom} of a FQHE
state (defined on lengthscales  large
compared to $\ell_B$), 
and not merely  a variational parameter.

In its finite-$N$ polynomial form (\ref{polyl}), the Laughlin state
$|\Psi^q_L(g)\rangle $ is an
eigenstate of  $L_R(g,0)$ where
$L_R(g,\bm r)$ = $g_{ab}\Lambda^{ab}(\bm r)$
 generates
rotations of the guiding-centers about a point $\bm r$; here
$\Lambda^{ab}(\bm r)$ = $\Lambda^{ba}(\bm r)$   are the three generators of
area-preserving linear deformations\cite{fdmharxiv} of the guiding-centers around $\bm r$: 
\begin{equation}
\Lambda^{ab}(\bm r) = \frac{1}{4\ell_B^2}\sum_i \{\delta
R^a_i(\bm r),\delta R^b_i(\bm r)\},
\end{equation}
with 
$\delta R^a_i(\bm r) \equiv R^a_i-\bm r$.
Leaving $\bm r$ implicit, these obey the non-compact Lie algebra\cite{fdmharxiv}
\begin{equation}
[\Lambda^{ab},\Lambda^{cd}] = -\frac{i}{2}\left (
\epsilon^{ac}\Lambda^{bd} +
\epsilon^{bd}\Lambda^{ac}
+ a \leftrightarrow b\right ),
\end{equation}
which is isomorphic to $SO(2,1)$, $SL(2,R)$, and $SU(1,1)$, 
with a Casimir $C_2$ =   $ -\frac{1}{2}\det \Lambda$ $\equiv$
$-\frac{1}{4}\epsilon_{ac}\epsilon_{bd}\Lambda^{ab}\Lambda^{cd}$.  
 
FQHE states with $\nu$ = $ p/q$ can be simply understood as
condensates of  ``composite bosons''\cite{gmpbos}
which are ``elementary droplets'' of the incompressible fluid
consisting of $p$ identical charge-$e$ particles ``bound to $q$ London  flux
quanta'' (\textit{i.e.}, occupying $q$ one-particle orbitals of the
Landau level), which behave as a boson under interchange.  This
requires that the Berry phase cancels any bare
statistical phase: 
$(-1)^{pq}$  =  $\xi^p$, where $\xi $ = $-1$  ($+1$)  for fermions (bosons). 
 For a condensate of charge-$pe$ objects, the elementary
 fractionally-charged vortex has charge $\pm e^*$ = $\pm (\nu e^2/h)
\times (h/pe)$ = $\pm e/q$.  This work aims to extend the description
of the ``composite boson'' by giving it (2D orbital) ``spin'' and geometry.

Polynomial FQHE wavefunctions like (\ref{polyl})
that describe $\bar N$ = $N/p$ = $N_{\Phi}/q$  elementary droplets are generically  eigenstates of
$L_R(g,0)$ with eigenvalue $\frac{1}{2}pq\bar N^2 + \bar s\bar N$,
where $\bar s$ is a variant of the so-called ``shift'' that I will
identify as a fundamental FQHE parameter, the \textit{guiding-center spin},
that characterizes the geometric degree of freedom  of FQHE states.
It can also be obtained as the limit $\bar N \rightarrow \infty$ of
\begin{equation}
\bar s = \frac{1}{\bar N}\sum_{m= 0}^{q\bar N -1}
(m+{\textstyle\frac{1}{2}})(n_m(\bar g,\bm r)-\nu),
\end{equation}
where  $n_m(\bar g,\bm r)$, $m \ge 0 $ are the occupations of
guiding-center
orbitals defined as the  eigenstates of $L_R(\bar g,\bm r)$.

Note that the ``superextensive'' ($\propto \bar N^2$) contribution to
the eigenvalue derives from the uniform background density
contribution $\nu \delta^2_{\bm q,0}$  to $\rho_{\bm q}$, 
and can be removed (regularized) by defining
$\Lambda^{ab}(r)$ in the thermodynamic limit $N_{\Phi} = q\bar N\rightarrow \infty$
using the limit of the $\bm q\ne 0$ $SU(N_{\phi})$ generators alone, which become
continuous functions $\rho(\bm q)$ of $\bm q$,  with $\lim_
{\lambda \rightarrow 0} \rho(\lambda \bm q)$ = 0. Then
\begin{equation}
\Lambda^{ab}(\bm r) = 
\lim_{\lambda  {\rightarrow 0}}  \left (-\frac{1}{2}\frac{1}{(\lambda\ell_B)^2}\frac {\partial}{\partial q_a}\frac {\partial}{\partial q_b}
\rho(\lambda \bm q)e^{-i\lambda \bm q\cdot \bm r}\right ).
\end{equation}
The Laughlin state $|\Psi^q_L(\bar g)\rangle$ is  an eigenstate 
of $\bar g_{ab}\Lambda^{ab}(\bm r)$
with $\bar s$ = $\frac{1}{2}(1-q)$.   Note that for fermionic particles ($\xi$ =
$-1$), $\bar s$ is odd under particle-hole transformations, and vanishes
when the Landau-level is completely filled (here $q$ = 1).   A
spin-statistics selection rule requires that
\begin{equation}
(-1)^{2\bar s}(-1)^{2s} = (-1)^{pq} = \xi^p, \quad (-1)^{2s} =
(-1)^p,
\end{equation}
where $s$ is the  (orbital) ``Landau-orbit spin'' of the elementary
droplet ($s$ = $-\frac{1}{2},-\frac{3}{2},\ldots$ for particles with
  Landau index $0,1,\ldots $).  The expression for $\bar s$ may now be
  inverted to define the  (local)
  unimodular guiding-center metric $\bar g_{ab}(\bm r)$  by the
expectation value
\begin{equation}
\lim_{\bar N \rightarrow \infty} \frac{1}{\bar N} \langle
\Psi^R|\Lambda^{ab}(\bm r)|\Psi^R\rangle  =
{\textstyle\frac{1}{2}}\bar s \bar g^{ab}(\bm r), \quad \det \bar g = 1,
\end{equation}
so if $\bar \rho(\bm r)$ is the local droplet density, $\frac{1}{2}\bar
s \bar \rho(\bm r) g^{ab}(\bm r)$ is the local density of the
deformation generator.

The quantization of $2\bar s$ as an integer is a topological property deriving from  the incompressibility
of  FQHE states.   A simple picture that is reminiscent of
Jain's notion of ``quasi-Landau-levels''\cite{jain} supports this:  the ``elementary
droplet'',
with a shape fixed by $\bar g_{ab}(\bm r)$, supports $q$
single-particle 
orbitals with guiding-center spins  $\frac{1}{2},\frac{3}{2},\ldots
,\frac{q-1}{2}$.   The way these are occupied by the $p$ particles of
the droplet, determines the guiding-center spin of the droplet as the
actual total guiding center spin of the configuration, minus that ($\frac{1}{2}pq$)
given by putting $p/q$ particles in each orbital.   The repulsive exchange and
correlation
fields of particles outside the droplet will give each of the internal
levels a mean energy for orbiting around an effective potential
minimum at  its center.   The droplet will be stable, with a quantized
guiding center spin that is adiabatically conserved as the droplet
changes shape, provided there is a finite positive energy gap between
the highest occupied and lowest empty single-particle state in the
droplet.    Collapse of this gap implies  that the system has become
compressible with an unquantized or indeterminate value of $\bar s$.

The geometrical degree of freedom exposed here also suggests a new
look at the problem of formulating a continuum description of
incompressible FQHE states.  
Elsewhere, I will present a continuum description combining Chern-Simons fields
with the geometry  field $\bar \omega_a(\bm r,t)$, where $\bar
g_{ab}$ = $\bar \omega^*_a\bar \omega_b + \bar \omega^*_b\bar \omega_a$,
but mention here some
fundamental
formulas that emerge.    First, the electric charge  density is given
by $pe\bar \rho(\bm r)$, where $\bar \rho(\bm r)$ is the local elementary
droplet
(composite boson) density, and
\begin{equation}
\bar \rho(\bm r,t) = \frac{1}{2\pi p q}\left ( \frac{ pe}{\hbar} B(\bm
  r) + \bar s K(\bm r,t)\right ),
\end{equation}
 Here $B(\bm r)$ is the 
externally-imposed 2D (normal)  magnetic flux
   density,
(assumed to be time-independent, but not necessarily spatially uniform), and
$K(\bm r, t)$ is the instantaneous Gaussian curvature of the
unimodular guiding-center-metric field $\bar g_{ab}(\bm r,t)$, given
by $K$ = $\epsilon^{ab}\partial_a\Omega^{\bar g}_b$, 
$\Omega^{\bar g}_a$ = $\epsilon^{bc}\bar \omega^*_b\nabla_a^{\bar
  g}\bar \omega_c$,
where $\Omega^{\bar g}_a$ is the spin connection  gauge-field and $\nabla^{\bar g}_a$
is the covariant derivative (Levi-Civita connection) of $\bar g_{ab}$.
 This formula
could perhaps have been anticipated from the work of Wen and
Zee\cite{wenzee}, who considered coupling  Chern-Simons fields to
curvature, but the curvature they apparently had in mind was not the
collective dynamical internal degree of freedom described here, but
that due to  placing the FQHE system on a curved 2D surface embedded
in 3D Euclidean space, as in formal calculations of the FQHE on a
sphere surrounding a monopole\cite{haldane83,halrez85}.   The second formula is that
the canonical conjugate of  the geometry field $\bar \omega_a(\bm r)$
is
\begin{equation}
\bar \pi^a_{\bar \omega}(\bm r) = \hbar \bar s \bar \rho(\bm r)
\epsilon^{ba}\bar \omega_b(\bm r)^*,
\end{equation}
so the momentum density (translation generator density)  is
$\bar \pi^b_{\bar \omega}\nabla_a^{\bar g}\bar \omega_b $ = $\hbar s
\bar \rho \Omega^{\bar g}_a$.   These formulas parallel those of
quantum Hall ferromagnets, with guiding-center spin and Gaussian
curvature replacing true electron spin and Berry curvature.    On
large lengthscales, the elementary charge $e^*$ = $\pm e/q$ quasiparticles appear
as rational cone-singularities of the metric field $\bar g_{ab}(\bm
r,t)$ with localized
Gaussian curvature $K$ = $\pm 4\pi/(2\bar s )$.

In summary, the prevalent assumption of rotational
invariance of FQHE fluids  conceals a fundamental geometric
degree of freedom, the shape of their correlations, described by a
unimodular spatial metric field that exhibits quantum dynamics.

This work was supported by DOE grant {DE}-{SC0002140}.
The author thanks the Laboratoire Pierre Aigrain, \'Ecole Normale Sup\'erieure, Paris, for its
hospitality during the final stages of this work.

\end{document}